\documentclass[multphys,vecphys]{svmult}
\usepackage{makeidx}         
\usepackage{graphicx}        
\usepackage{multicol}        
\usepackage[bottom]{footmisc}
\usepackage{amsmath}
\usepackage{amsfonts}
\usepackage{amssymb}
\usepackage{epsfig}
\usepackage{dcolumn}
\usepackage{bm}
\def\arcdeg{\hbox{$^\circ$}}
\def\nuc#1#2{\relax\ifmmode{}^{#1}{\protect\text{#2}}\else${}^{#1}$#2\fi}

\makeindex             

\def\kappaezz{(\tilde{\kappa}_{e-})^{ZZ}}

\begin{document}

\title{A precision test of the isotropy of the speed of light using rotating cryogenic
optical cavities}
\titlerunning{A precision test of speed of light isotropy}
\author{S. Schiller\and P. Antonini\and M. Okhapkin}
\institute{ \texttt{Institut f{\"u}r Experimentalphysik,
Heinrich-Heine-Universit{\"a}t D{\"u}sseldorf, 40225 D\"usseldorf, Germany.}
}
%
%

\maketitle
\begin{abstract}
A test of Lorentz invariance for electromagnetic waves was performed by comparing the  resonance frequencies of two stable
optical resonators as a function of orientation in space. The crystalline resonators were operated at 3.4\,K in a cryostat employing a pulse-tube 
refrigerator. A new analysis yields the Robertson-Mansouri-Sexl theory parameter combination
$\beta-\delta-1/2=(-0.6\pm 2.1\pm1.2)\cdot10^{-10}$ and one parameter of the Standard Model Extension theory,
$(\tilde{\kappa}_{e-})^{ZZ} = (-2.9\pm2.2)\cdot 10^{-14}$.
\end{abstract}

\section{Introduction}

The isotropy of space is a well-tested symmetry of nature \cite{bri79}. Because
it is a foundation of today's accepted theories of the fundamental
forces it continues to be the focus of both theoretical and
experimental studies. A series of experiments
\cite{lip03,mue03b,wol04a} have recently been performed with the goal of improving the limits for a hypothetical violation. They were in part
motivated by the development of an extension of the Standard Model (SME)
by Kosteleck\'{y} and coworkers \cite{Kos02hh,Kos01mb} that describes Lorentz violation in a comprehensive way.
This dynamical test theory indicates that isotropy violation, if it exists, may
exhibit characteristics that differ from those of previous kinematic
test theories, such as the Robertson-Mansouri-Sexl (RMS) theory \cite{man77}.

Here, we will not discuss the conceptual frameworks used to describe hypothetical violations of isotropy, 
since this is reported in the literature and is also treated in this Proceedings volume.
In this contribution, we limit ourselves to the description of an experiment used to perform an improved test
 of the isotropy of the speed of light. The experiment has already been presented previously
\cite{ant05}; here we give a more detailed description and report an extension of the data analysis.

The experiment was conceived as an actively rotated Michelson-Morley
experiment using ultrastable optical cavities interrogated by lasers. It was a
natural extension of our previous work with stationary resonators
\cite{mue03b,bra02}, but employed a completely new apparatus, except
for the sapphire optical cavities.

The experiment consists in measuring the  difference (beat) $\nu_1-\nu_2$ between the  frequencies 
of two longitudinal modes of two orthogonal standing-wave cavities as a function of orientation
in space.  If isotropy is violated, according to the Robertson-Mansouri-Sexl test theory and the Standard Model extension,
the beat frequency will vary as
\begin{equation}
\frac{\delta(\nu_1(t)-\nu_2(t))}{\nu}=
2 B(t) \sin{2\theta(t)} +2 C(t) \cos{2\theta(t)},\label{eq1}
\end{equation}
where $\nu_1\approx\nu_2\approx\nu$ is the
average frequency ($2.8\cdot 10^{14}$\,Hz in this experiment) and $\theta(t)$ is the angle between one cavity's
axis relative to the south direction. 

In the RMS test theory,
the amplitudes 2$B(t)$ and 2$C(t)$ are proportional to the parameter
combination $\beta-\delta-1/2$, where $\beta,\,\delta$ parametrize
deviations of the frame transformation equations from the usual
Lorentz form. The explicit form is given below.

 In the SME test theory \cite{Kos02hh}, each amplitude 2$B(t)$ and 2$C(t)$ is a linear combination of eight
coefficients weighted by time-harmonic factors. The amplitude $B(t)$ contains frequency components at
0, $\omega_{\oplus}$, $2\omega_{\oplus}$, $\omega_{\oplus}\pm\Omega_{\oplus}$ and
$2\omega_{\oplus}\pm\Omega_{\oplus}$,
while $C(t)$ contains in addition one component at the frequency $\Omega_{\oplus}$. Here $\omega_{\oplus}$ is
Earth's sidereal angular frequency and $\Omega_{\oplus}$ is Earth's orbital frequency.
The determination of the individual $\tilde{\kappa}_{o+}$ coefficients requires the ability to
resolve the contribution of Earth's orbital motion in order to discriminate between modulation frequencies
differing by $\Omega_{\oplus}$. Thus a measurement extending over at least 1 year is necessary.
However, in this experiment we concentrated on a single parameter, $\kappaezz$, and using previous results for the remaining parameters, it was
possible to obtain a result within a much shorter measurement time.

\section{Experimental setup}

\begin{figure}[h]
\centerline{ \epsfysize=9cm\epsffile{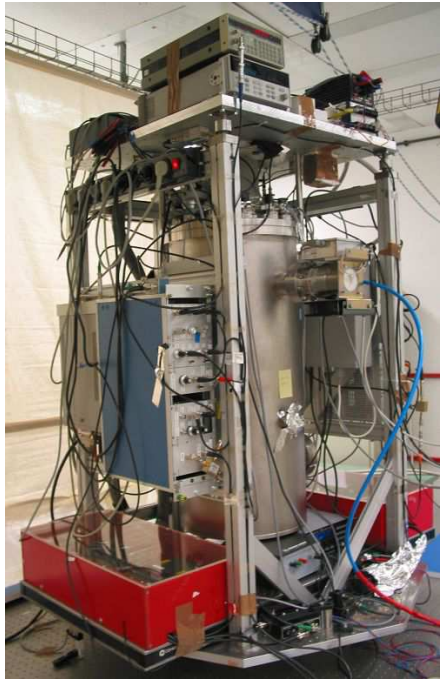} }
\caption{\small The experimental setup. 
The turntable is located under the octagonal base plate 
and is not visible in the photograph. See text for details. \label{fg:set-up}}
\end{figure}

An overall view of the experimental setup is shown in Fig.\,\ref{fg:set-up}.
The whole setup was actively rotated by a computer-controlled precision ball bearing turntable. The
turntable itself rested on an optical table (3\,m $\times$ 1.5\,m) that was not floated.
An octagonal base plate mounted on the turntable carried most of the components, except for the vacuum forepump, the Helium
compressor, and a synthesizer.
The cryostat was attached via columns to a plate that could be rested on top of a rack mounted on the base plate. 
Cryostat and plate could be removed from the rack for opening.
On the plate the rotary valve and its driver, temperature
controllers, dataloggers, synthesizers and power supplies were
installed. The servo systems for the cavity frequency locks and the
laser power stabilizations were mounted on the sides of the rack. The
laser systems, enclosed by boxes, were located on a breadboard
placed on the octagonal base plate. The beat frequency detector was
contained in one of the boxes.
Thermal insulation (not shown in the figure) was used to shield several of the components of the setup.

\subsection{The cryogenics}

\begin{figure}[t]
\centerline{\epsfysize=8cm \epsffile{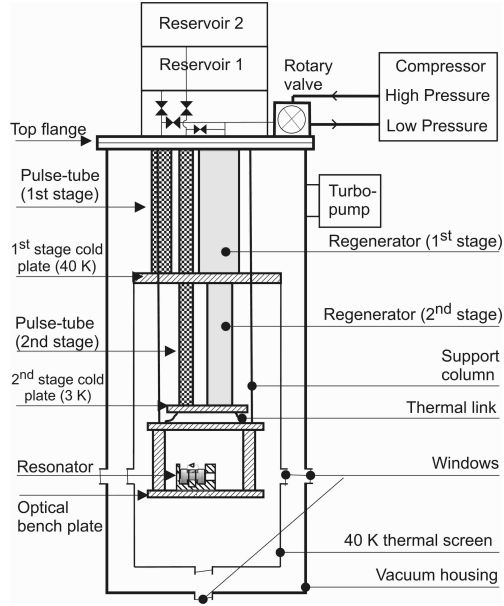}              }
\caption{\small Cut-away schematic of the cryostat with internal components (not to scale). Height and diameter of the vacuum housing are 90 and 40\,cm, respectively.
 Only one resonator is shown. }
 \label{fg:prk}
 \end{figure}

The cryostat is sketched in Fig.\,\ref{fg:prk}.
Cooling was implemented by a two-stage pulse-tube cooler \cite{wan97a} (Transmit
GmbH), using high pressure (18 bar) Helium as working medium.
The first stage had a high cooling power (6\,W at 50\,K) and
reached a temperature of approx.\,41\,K, while the second had a
lower cooling power (approx.\,0.2\,W at 3\,K) but was able to reach a
minimum temperature of about 2.2\,K without load, and 3.2\,K when
loaded with the experimental set-up used for this experiment. The
cooler was driven by a water-cooled Helium compressor (Leybold) with
6\,kW power consumption. A significant advantage of this novel
cooler type compared to standard cryocoolers is the absence of moving parts inside the cryostat; only
the He gas moves within the cryostat, under periodically modulated pressure. Mechanical motion was, however, present
in the rotary valve on the top plate. As a consequency, the displacement of the resonators was modulated at the rotary 
valve frequency, with an amplitude of approx.\,1\,$\mu$m vertically and horizontally, as determined from the propagation of the laser beam
exiting the cavities.

An optical cryostat was used, containing three free-space optical
access ports: two half-inch diameter windows for horizontal access
and an additional window located at the bottom (not used here).
Anti-reflection-coated BK7 was used as window material. To avoid backreflections, the windows' normals
are angled with respect to the beam direction. The lateral ports
were used during the laser alignment and frequency lock phases to identify the
cavity modes excited by the lasers. 

A copper heat shield (thermal screen) was attached to the
first stage cold plate. To improve shielding from the 300\,K vacuum
can, it also contained angled windows. The space available below the
second stage cold plate was about 30\,cm in height and 30\,cm in
diameter. The experimental assembly consisted of an upper plate,
four columns, and a bottom optical bench plate, all made of copper, rigidly connected together. 
It was attached to the top flange by
3 hollow stainless steel rods of 40\,cm length, which were heat sunk to the first stage cold plate but not to
the second stage cold plate \cite{lie01}. Copper mesh provided the
thermal link between the experimental assembly and the second stage
cold plate. The assembly was thermally shielded by superinsulation foils.

A turbo pump was used to continously evacuate the chamber. The top
flange of the cryostat contained twelve KF flanges that were used for
electrical signal and optical fiber vacuum feedthroughs.

After evacuation of the cryostat, the cool-down time to 3.2\,K was
about 12 hours.

On a timescale of 10\,min, the temperature of the first stage varied
by less than 0.1\,K. On the optical bench plate, where the cavities are located, the variations were significantly lower, with a 4~mK
short-time (10~s) temperature instability 
and 50~mK instability over long times (10\,h). These variations came from (small) instabilities of the
pulse-tube cooler itself, and from a dependence of the temperature
of the cold stages on the ambient temperature. To reduce these
variations, the temperature of the resonators was kept stable at a value slightly above
the second stage temperature by
active temperature control. A combination
of a heater attached to the underside of the optical bench plate, equidistant from the two optical
resonators, and a high sensitivity thin film sensor fixed to one
resonator housing, together with a
 commercial digital temperature controller was used, 
and kept the temperature constant at 3.4\,K.

\subsection{The optics setup}

\begin{figure}[t]
\centerline{ \epsfxsize=9cm\epsffile{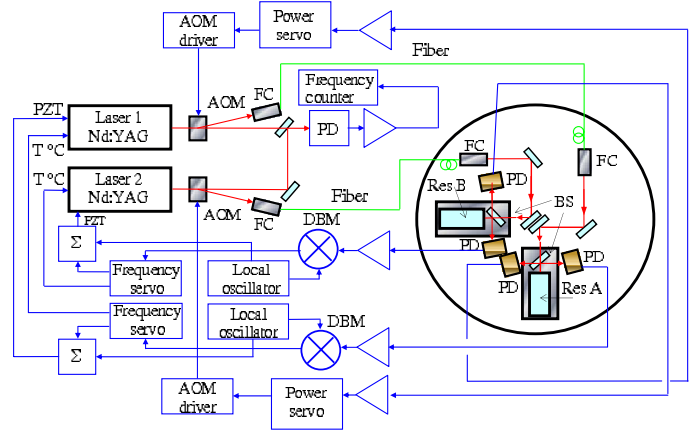} }
\caption{\small
 Schematic of the opto-electronic system. Two Nd:YAG lasers (1064\,nm)
are frequency-locked to two sapphire optical resonators
located in the cryostat. The beams are fed to the resonators via optical fibers.
Acousto-optic modulators (AOM), stabilize the power of the beams fed to the resonators.
DBM: doubly-balanced mixer; BS: beam splitter; PD: photodiode; FC: fiber coupler, PZT:
piezoelectric frequency control actuator; T: temperature control of the
laser crystal. }
\label{fg:setupfreqlock}
  \end{figure}

\begin{figure}[t]
\centerline{ \epsfxsize=8cm \epsffile{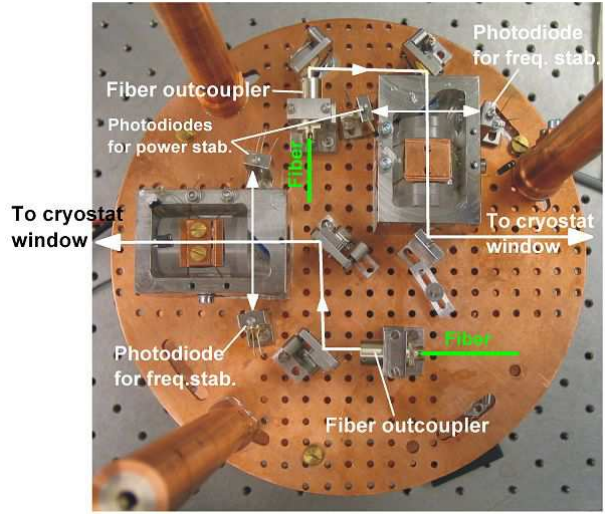}                }
\caption{\small View of the main cryogenic optical components located on the optics bench plate, 
without electrical cables and optical fibers. The resonator housings shown were later replaced by 
gold-plated copper housings. 
\label{fg:set-up02}}
  \end{figure}

An overview of the functional parts of the laser and resonator system is shown in Fig.~\ref{fg:setupfreqlock}, and
the components on the optics base plate are shown in Fig.\,\ref{fg:set-up02}.

The two optical resonators are made of pure sapphire (Al$_2$O$_3$)
\cite{see97a}. Each consists of a 3\,cm long cylindrical spacer with
inner diameter 1.0\,cm, outer diameter 2.6\,cm,  and crystal c-axis
parallel to the cylinder axis. The sapphire mirrors have 1\,m radius
of curvature and are optically contacted to the spacer. A small hole
perpendicular to the cavity axis serves for evacuation. The mirrors
are coated for high reflection at 1064~nm. These resonators were
already used for relativity tests \cite{mue03b,bra02,sto98a,mue02}.
The linewidths of the two resonators were 100\,kHz at the time of
the experiment.

Sapphire was chosen because of its very low thermal expansion
coefficient at cryogenic temperature and its low dimensional drift
\cite{sto98a}. For dielectric crystals, the thermal expansion is due solely to phonons and therefore the
expansion coefficient drops as $T^3$ as the temperature $T$ approaches zero. For
sapphire the value at 3.4~K  is approx.\,$8\cdot 10^{-11}\, \mathrm{K}^{-1}$ \cite{luc03}.
We note, however, that the effective expansion coefficient of a mounted cavity can differ
substantially, because it also involves the interaction of the resonator with its
holder. Thus, in past experiments we have found zero crossings of the thermal expansion
coefficient at 3\,K \cite{sto98-thesis}.

In the present setup, the two resonators were mounted in two
housings made of invar and coated with a
5~$\mu$m thick gold layer. Gold was used to obtain good contact
between the resonators and their housings. The resonators were
then fixed to the housing using thin copper straps. The
straps were not strongly tightened, to avoid squeezing the
resonators because of contraction of the straps during cool-down.

The laser beams were transported to the
cavities via two 4\,m long polarization-maintaining single-mode fibers
with 8$\arcdeg$ angled fiber ends (to avoid back reflections and miminize etalon effects). The fiber ends were
connected to fiber couplers containing short focal length lenses and rigidly attached to the optical
base plate. For each laser beam, two adjustable vacuum compatible
stainless steel mirror mounts deflected the light toward a 90\% reflection/10\% transmission beam 
splitter placed just before the cavity. The light
directly reflected from the beam splitter reached a 2\,mm diameter
InGaAs photodiode that provided the signal for power monitoring and stabilization. 
The transmitted light was reflected
by the cavity, then partially by the beam splitter and was sent to another photodetector
of the same type to provide the signal for frequency locking.

The alignment of the laser beams was done at room temperature with
the cryostat open. After cooling, the lasers were locked to the
TEM$_{00}$ modes, which could be identified with the help of CCD
cameras that monitored the cavity light leaving the cryostat through
the windows. The coupling efficiency was typically 10\% at room
temperature. The efficiency was reduced by a factor 2 in the cold state due to differential thermal contraction
effects.

Two diode-pumped monolithic non-planar ring oscillator Nd:YAG lasers
emitting 200\,mW at 1064 nm were used. The power fed into the fibers
was about 1.5\,mW. The Drever-Hall reflection locking scheme was
used \cite{dre83a}. The lasers were phase modulated at frequencies of 300\,kHz via the piezoelectric actuators acting on the
laser crystals \cite{can94}. The error signals had typical SNR
$> 10$ in a bandwidth of 100\,kHz after amplification. They were
processed by respective analog servos, each employing 
a loop for the laser piezoelectric actuator (unity gain coefficient at about
15\,kHz) and a slow loop for laser crystal temperature control.
The accuracy of the servo electronics was better than 0.1\,Hz at 100\,s integration time.

The power of each laser beam incident on the cavities was about 50 to 100\,$\mu$W and was 
 actively stabilized to a relative level of $1 \cdot 10^{-4}$ using an
acousto-optical modulator (AOM) placed before the fiber outside the cryostat. 
The AOMs also served as optical isolators.

On the laser breadboard, two parts of the laser beams were superimposed on a fast photodiode,
producing a heterodyne signal at the beat frequency $\nu_1 -
\nu_2$, about 700~MHz. This frequency was
mixed down with a synthesizer to a frequency of about
10~MHz, to exploit the higher accuracy of the counter at lower
frequencies. Both synthesizer and frequency counter were phase-locked to the 5 MHz output of a
hydrogen maser.

\section{Characterization of the setup}

After cool-down, the beat frequency initially exhibited a drift on
the order of 1\,Hz/s. After 2 months, this was reduced to
0.02\,Hz/s. Our following discussion refers to this stable regime of
operation.

\begin{figure}[t]
\centerline{ \epsfxsize=7cm \epsffile{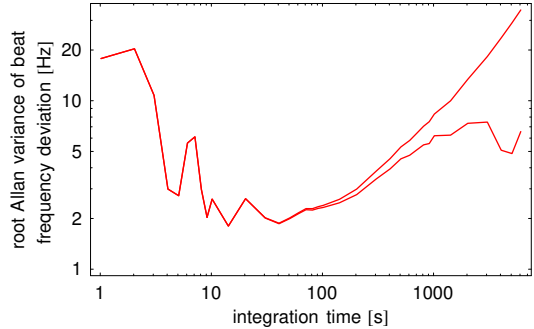}}
\caption[]{\small Root Allan variance of beat frequency under rotation (without tilt correction 
and temperature decorrelation). 
Upper curve: from raw data including drift;
lower curve: after removal of the drift (approx.\,0.02\,Hz/s) from raw data. 
The time record is the same as in Fig.\,\ref{fg:decorr}, left.}
\label{fg:RAV}
\end{figure}

A main characteristic of the apparatus, the frequency instability of
the beat, is shown in Fig.\,\ref{fg:RAV}. The root Allan variance
(RAV) exhibits a peak at about 7\,s. This is due to the modulation
of the beat frequency by the pulse-tube cooler with peak-peak 
amplitude of about 400\,Hz. Although the cooler 
has a mechanical frequency at 1.1\,Hz, this frequency was aliased to
a lower frequency by the 1\,s sampling time of the frequency
counter. 
The minimum RAV ($7 \cdot 10^{-15}$) is attained at $\tau=20 - 30$\,s. At the
half-period of the rotation, 300\,s, the RAV has increased to $1.4 \cdot 10^{-14}$. 
The main reasons for this level appear
to have been due to a low signal-to-noise ratio of the error
signal and to the presence of mechanical/thermal noise from the cooler components and from the laboratory. 
These RAV values are about an order of magnitude
higher than those obtained for the same integration time in the previous, nonrotating, experiment 
using these resonators \cite{mue03b}. However, for the purpose of the isotropy test the 
relevant integration time was $\tau=12\,$h  in the latter
experiment, for which the instability was higher than that  at 300\,s, relevant in the present experiment.

\subsection{Laser power }

The dependence of the TEM$_{00}$ resonance frequencies on the power
of the beam impinging on the resonators could reliably be measured only for relatively large power 
changes ($>5\,\mu$W) and was dependent on the mode-match efficiency. For the resonator with the larger sensitivity,
a conservative upper limit is 50\,Hz/$\mu$W. 
For the power levels
reaching the cavities and the relative power instability given above,
this implies an influence of the residual power fluctuations of
not more than 0.5\,Hz (1.8 $\cdot$10$^{-15}$).

\subsection{Tilt of the resonators}\label{sect:tilt}

The sensitivity of the beat frequency on the orientation of the cryostat was measured. 

The experiment was operated on an optical table that was not floated but was supported by a metal
frame. The lengths of the 
 feet of the frame could be changed by acting on screws.
Before each run, the tilt sensitivities were determined by measuring the
 frequency shift as function of the inclination  of the cryostat in two orthogonal
directions. 
 To this end,
the leg screws were turned and the orientation of the optical table
as a whole was changed. The resulting cryostat inclination was
measured by a sensitive two-axis tilt-sensor attached to the plate that holds the cryostat. The resolution of the
tilt sensor was 0.1~$\mu$rad. The tilt sensor output showed a small
dependence on the temperature, but this effect was suppressed by a
passive temperature insulation.

We measured a sensitivity of about 0.06~Hz/$\mu$rad for tilts around two axes
parallel to the resonator axes.
Typically, the peak-to-peak tilt variation during rotation was 80$\,\mu$rad, corresponding to
5\,Hz beat modulation. Because of this magnitude, the tilt effects were taken into account in the data analysis.

\subsection{Temperature sensitivities}

We measured the dependence of the beat frequency on the temperature of the optical bench
 by changing the set-point of the temperature controller. The temperature was measured using a
Cernox  sensor (Lakeshore), connected to the housing of one of the two
cavities. 
Because of the high thermal conductivity of the bench the temperature difference between the two resonators
is expected to be very small.
The temperature sensitivity was 1.5~Hz/mK. This is equivalent to a  thermal expansion coefficient difference between the
two cavities of $5.3\cdot10^{-12}$/K, a value 15 times lower than the nominal expansion coefficient of sapphire at
the same temperature.

\begin{figure}[h]
\centerline{ \epsfxsize=8cm \epsffile{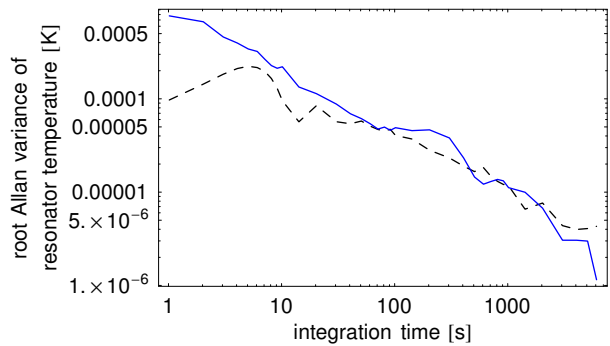} }
\caption{\small Root Allan variance of the (actively stabilized) temperature measured near one of the
resonators. Full line: with rotations: the plateau at 300\,s is due to a (residual) modulation correlated with rotation.
The time record is the same as in Fig.\,\ref{fg:decorr}.
Dashed line: from a record without rotations.} \label{fg:AV-texp}
\end{figure}

The typical residual temperature variation correlated with rotation are 0.15 $\,$mK peak-to-peak,
leading to an influence on the beat frequency of 0.2\,Hz ($0.8\cdot10^{-15}$).
The instability of the temperature of the optical bench for various time scales is shown in
Fig.\,\ref{fg:AV-texp}. 

Ambient and cryostat component temperatures did have a significant effect on the beat frequency.

The sensitivity of the beat frequency to changes of the temperature
in the lab was measured by acting on the air conditioning system. A
sensitivity of approx.\,75~Hz/K was measured.
We did not observe any strong correlation
between the lab temperature and the temperature of the optical bench cold plate. 
When the lab temperature was modulated so as to give 40\,Hz\ beat modulation, 
the cold plate temperature's
peak-peak amplitude was smaller than 0.3\,mK, corresponding to a calculated
beat frequency modulation of less than 0.5\,Hz.
Thus, it appears that the observed temperature sensitivity of the beat
frequency was due to the thermal sensitivity of the fibers. 
Temperature changes affected their optical path length and in the presence of 
residual amplitude modulation and spurious etalons
the laser frequency lock point changed. The fiber temperature
is influenced both directly by the ambient temperature (detectable by heating the fibers locally) and by their contact to
various parts of the cooler inside the cryostat, whose temperatures
also changed with ambient temperature and cooler operating
conditions.

Typical temperature variations measured at the top of the cryostat during one rotation were as low as 25 mK peak-peak
after improvements in the temperature insulation of the setup, corresponding to a 2\,Hz peak-to-peak effect assuming
the above temperature sensitivity.

\begin{figure}[h]
\centerline{\epsfxsize=6cm\epsffile{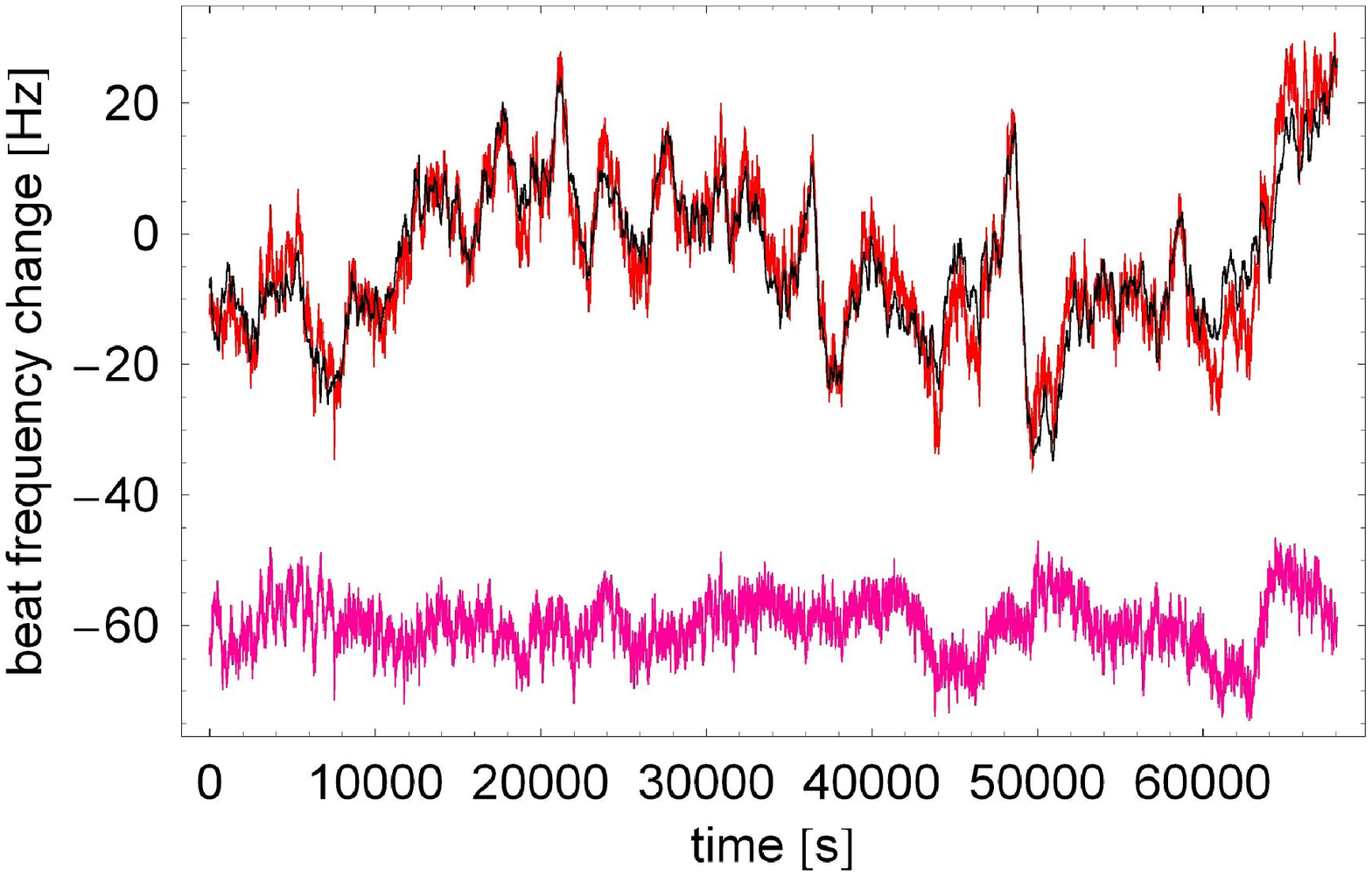}
\hskip-.1cm
\epsfxsize=6cm\epsffile{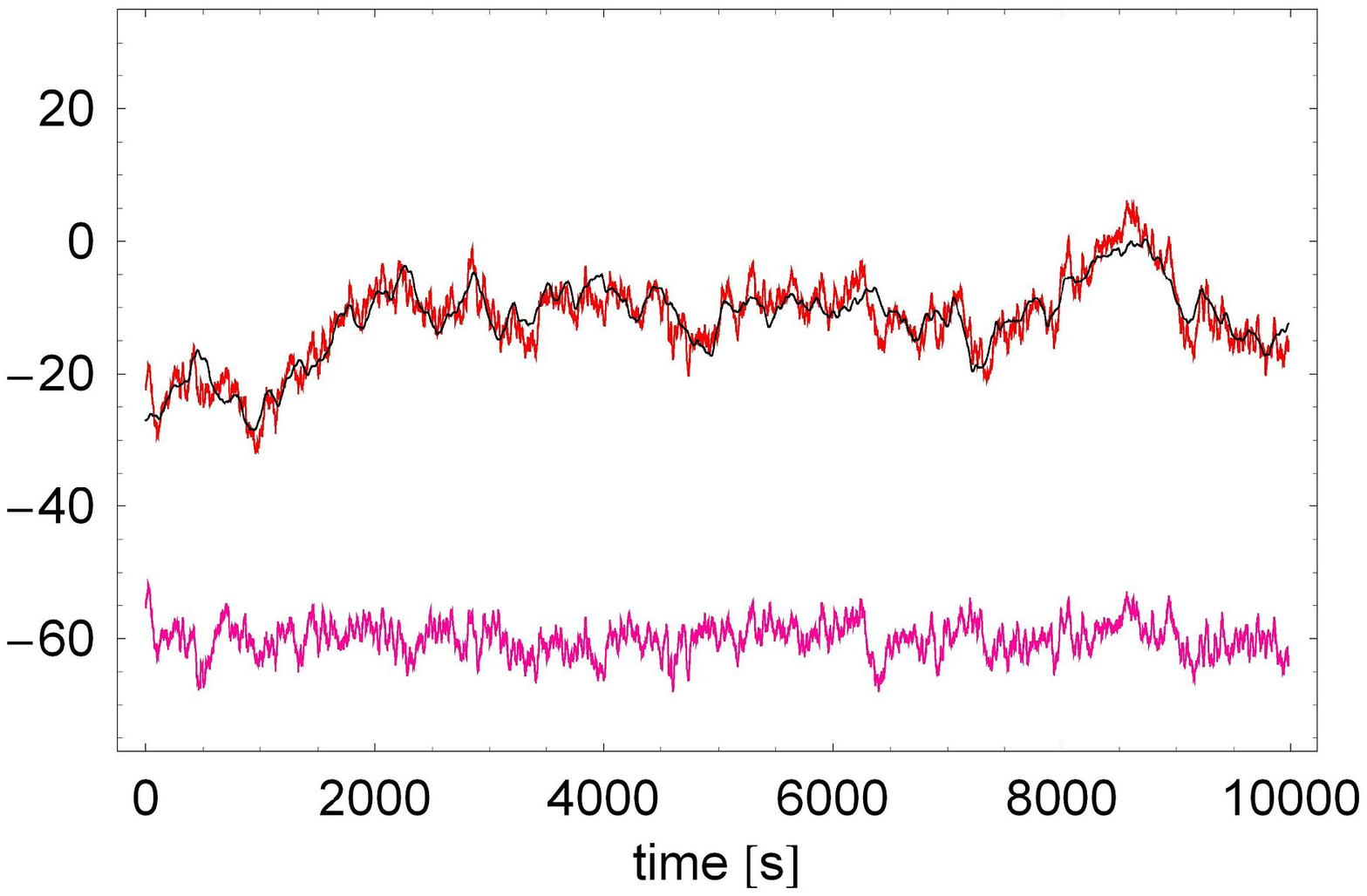}
}
\caption{Influence of temperature variations. Left: decorrelation of a 70\,ks long beat frequency data set; 
right: decorrelation of a 10\,ks subset starting at 50\,ks. 
Red: beat frequency, after subtraction of linear and quadratic drift and correction for tilts. 
Black: Linear combination of seven temperature traces that
best fits the corrected beat frequency. 
Pink (bottom traces): residuals, offset for clarity.
During this measurement, the apparatus was rotated. Traces are averaged over 21 s.}
\label{fg:decorr}
\end{figure}

In order to characterize to what extent the various temperature variations induced beat frequency variations,
we analyzed the correlations. The monitored temperatures were two laboratory temperatures measured
at the top and bottom of the cryostat, four temperatures on different cooler components inside the cryostat and the temperature of one resonator holder.
A linear regression analysis showed that there are strong correlations between these parameters and the beat frequency.
Fig.\,\ref{fg:decorr} shows an example of this analysis.
As can be seen, the slow variations of the beat frequency with respect to a nearly constant drift are to a large extent
removed.


A summary of the systematic effects is given in Table~\ref{TableSyst}. Not included are sensitivities to the temperatures
of individual refrigerator components, since they could not be measured.

\begin{table}[b]
\caption{\small Characterized systematic effects. The column "Inst/mod" lists the measured parameter instability
on the timescale of the rotation half-period or the peak-peak modulation. The column "Systematic"
is the product of the first column and the sensitivity coefficient.
The estimates of the systematics are indicative, since we do not distinguish between angular variations of 
type $cos\theta$, $sin\theta$ and $cos2\theta$, $sin2\theta$,  
the two latter types being the relevant ones for the isotropy test.}
\centerline{
\begin{tabular}{lllll}
\hline
\noalign{\smallskip}
Effect & Inst/mod\ & Sensitivity\ & Systematic\ & Relative \\
\noalign{\smallskip}\hline\noalign{\smallskip}
Tilt & 50 $\mu$rad & 0.06 Hz/$\mu$rad & 3 Hz & 1.1 $\cdot$ 10$^{-14}$ \\
Ambient temperature & 0.025 K & 75 Hz/K & 2 Hz & 0.7 $\cdot$ 10$^{-14}$ \\
Resonator temperature & 150 $\mu$K & 1.5 Hz/mK & 0.2 Hz & 0.8 $\cdot$ 10$^{-15}$ \\
Laser power  & 10 nW & 50 Hz/$\mu$W & 0.5 Hz & 1.8 $\cdot$ 10$^{-15}$ \\
\noalign{\smallskip}
\hline
\end{tabular}
}
\label{TableSyst}
\end{table}


\section{Data collection and analysis}

A computer-controlled rotation stage rotated the cryostat over a range of 90$^\circ$.
The total range accessible was limited to a value slightly  above this by the He pressure lines connecting
the pulse-tube cooler with the stationary compressor. The period of rotation was chosen as 600\,s.
Shorter periods led to a significant shaking of the cryostat and were therefore not used.

Two synchronized computers collected the data. 
One computer controlled the rotation angle and rotation speed of the experiment, recorded the beat frequency and the temperature
of the optical resonators, by means of a program written in LabView. Sampling time was 1\,s.
The second computer recorded the cryostat tilt angles, and several temperatures as mentioned above.
The tilt and temperature values were used in the data analysis as explained below.

The data discussed here was obtained after about two years of
testing and improvements on the whole system. Test runs performed
initially typically exhibited significant drifts of
properties of the apparatus, such as cooler
internal temperatures. After minimizing these variations, we
succeeded in obtaining stable operation.


For data analysis, the (aliased) beat frequency modulation was
removed from the data by filtering in the Fourier domain, and the
sum of the tilt angles multiplied by the respective tilt
sensitivities was subtracted. Decorrelation of the temperatures was then performed, if desired.

\begin{figure}[t]
\centerline{
\epsfxsize=8cm
\epsffile{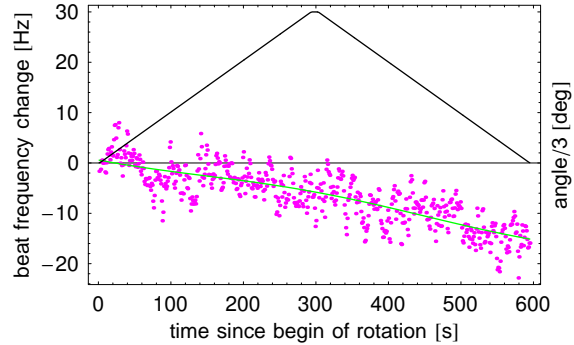}
}
\caption{\small
Beat frequency as a function of rotation of the apparatus.
Pink dots: tilt-corrected beat frequency, cooler-induced modulation removed, no averaging, 1\,s sampling time;
green line: fit of beat frequency according to Eq.(\ref{eq:rotfit}) with values 
$2C\nu=-0.76\pm0.18\,\hbox{\rm Hz}, 2B\nu=-0.02\pm0.41\,\hbox{\rm Hz}$;
black line: angular position.} 
\label{fg:beatclean}
\end{figure}

The beat frequency in each interval $\theta=[0^\circ;90^\circ;0^\circ]$ (labeled by $i$) was
least-squares fitted with the three-parameter function
\begin{equation}
a_i t + 2 B(t_i) \sin 2\theta(t)+2 C(t_i) \cos 2\theta(t),
\label{eq:rotfit}
\end{equation}
where the coefficient $a_i$ quantifíes a linear drift that may vary from rotation to rotation.
An example of an analysis of a single rotation is shown in Fig.~\ref{fg:beatclean}.
The obtained amplitude sets $\{2B(t_i)\}$, $ \{2C(t_i)\}$ are then analyzed according to the
Robertson-Mansouri-Sexl test theory and the Standard Model Extension.

\subsection{Analysis in the RMS framework}

It can be shown \cite{ant05} that according to the RMS test theory,
the amplitudes of the beat frequency modulation with angle are given
by

\begin{equation}\label{2b}
2B(t)=(1/2-\beta+\delta)({v^2}/{c^{2}_0})
(\gamma_3 \cos \omega_{\oplus}T_{\oplus}+
\end{equation}
\begin{displaymath}
\qquad \qquad
\gamma_4 \cos 2\omega_{\oplus}T_{\oplus}+\sigma_3 \sin\omega_{\oplus}T_{\oplus}
+\sigma_4 \sin 2\omega_{\oplus}T_{\oplus})\ ,
\end{displaymath}

\begin{equation}\label{2c}
2C(t)=(1/2-\beta+\delta)({v^2}/{c^{2}_0})
(\gamma_0+\gamma_1 \cos \omega_{\oplus}T_{\oplus}+
\end{equation}
\begin{displaymath}
\qquad \qquad
\gamma_2 \cos 2\omega_{\oplus}T_{\oplus}+\sigma_1 \sin\omega_{\oplus}T_{\oplus}+
      \sigma_2 \sin 2\omega_{\oplus}T_{\oplus}),
\end{displaymath}
where the constants are defined in Table~\ref{gamma-i}.

\begin{table}[b]
\caption{\small The values of $\gamma_i$ and $\sigma_i$
appearing in Eqs.(\ref{2b}) and (\ref{2c}).}\label{gamma-i}
\begin{center}
\begin{tabular}{clcl}
  \hline
  $\gamma_0$ = & $\frac{1}{4}\sin^2 \chi (3\cos^2 \Theta - 1)$ &  &   \\
  $\gamma_1$ = & $\frac{1}{2}\cos \Phi \sin 2\Theta \sin2\chi$ & $\sigma_1 $= & $\gamma_1\tan \Phi $ \\
  $\gamma_2$ = & $\frac{1}{4}\cos 2\Phi \cos^2\Theta(\cos 2\chi - 3)$ & $\sigma_2$ = & $\gamma_2\tan 2\Phi $\\
  $\gamma_3$ = & $\sigma_3\tan\Phi$ & $\sigma_3$ = & $\cos\Phi\sin\chi \sin 2\Theta$ \\
  $\gamma_4$ = & $-\sigma_4 \tan 2\Phi$ & $\sigma_4$ = & $\cos^2 \Theta \cos\chi \cos2\Phi$ \\
  \hline
  \end{tabular}
\end{center}
\end{table}

$T_{\oplus}$ is the time since the beginning of the data plus an offset that accounts for a time difference since the
coincidence of the lab's $y$ axis with the $\hat{Y}$ axis of the Sun-centered system \cite{Kos02hh}. The $y$ axis is parallel
to one cavity axis when in the $0^\circ$ position.
The direction of  the Sun's
velocity $\vec v$ relative to the cosmic microwave background is given by
the right ascension $\Phi=168^\circ$ and the declination $\Theta=-6^\circ$.

In order to provide an analysis similar to that of \cite{sta05,her05}, we consider
 a measurement interval extending over 183 hours that contained 940 rotations (after removal of a small number of outliers), grouped in 5 sets. 
Only the tilt correction was implemented and each set 
was fitted  with the functions (\ref{2b}) and (\ref{2c}) plus additional contributions $b_{syst}+ b'_{syst}t_i$ and $c_{syst}+ c'_{syst}t_i$ that
 model systematic effects. The individual sets yield the following values and standard errors for $\beta-\delta-1/2$:
 $\{(5.5, 5.0), (7.2, 9.0), (-8.3, 3.7), (5.7, 4.5),$ $(-2.6, 3.9)\}\cdot10^{-10}$.
The overall result is
\begin{equation}
\beta-\delta-1/2=(-0.6\pm 2.1\pm1.2)\cdot10^{-10}\ \ ,
\end{equation}
where the first error is statistical and the second error reflects the uncertainty in the experimentally determined tilt 
sensitivities and an estimate of the influence of laser power variations.

\subsection{SME test theory}\label{sme-results}\label{sme-analysis}

Considering that the time span over which data was taken was
significantly less than one year, the main goal of the data analysis
within the SME model was the determination of a value for
$\kappaezz$. The results of the cryogenic microwave experiment
\cite{wol04a} found the elements of $(\tilde{\kappa}_{e-})$ (except
for $(\tilde{\kappa}_{e-})^{ZZ})$ and the elements of $\beta_\oplus(\tilde{\kappa}_{o+})$
 to be at most several parts in
$10^{-15}$ in magnitude. If we assume these elements to be zero, we can use 
the result for $2C$ only to determine $(\tilde{\kappa}_{e-})^{ZZ}$,
\begin{equation}
(\tilde{\kappa}_{e-})^{ZZ}={4 \langle2C\rangle \over 3 \sin^{2}{\chi}}\ .
\end{equation}
For the set of rotations under the most stable conditions (of 76 hours duration, also
considered in \cite{ant05}) the average is $\langle
2C\nu\rangle=-2.4\,$Hz with a sample standard deviation of 1.9\,Hz,
and $\langle 2B\nu\rangle=0.8\,$Hz, with a sample standard deviation
2.6\,Hz. In this analysis, the frequency data was also decorrelated
with respect to the temperature data, whereby data intervals of
10\,ks were decorrelated individually, in order to allow for
changing environmental conditions. Without decorrelation of the
temperatures the values are, in the same order, ($-3.3$\,Hz, 2.3\,Hz) and
(2.8\,Hz, 2.4\,Hz) \cite{ant05}.

We used the large number of rotations performed under different experimental conditions to estimate the
uncertainty in $2C$ due to (identifiable) systematic effects at 1.8\,Hz. This includes
the  errors due to the uncertainty of the tilt coefficients and due to laser power variations. 
This results in\footnote{In \cite{ant05}, a factor 2 was inadvertently omitted when calculating $\kappaezz$ from $2C$.}
\begin{equation}
\kappaezz=(-2.9\pm2.2)\cdot10^{-14}\ \ ,
\end{equation}
where the uncertainty is dominated by the systematic effects.

\section{Conclusions}

The experiment described in this work was performed in order to improve the previous rotating laser 
experiment by
Brillet and Hall, exploiting some of advances in laser stabilization techniques developed since. Our experiment
yielded a significant improvement. Similar to their experiment, 
a strong limit for  $\beta-\delta-1/2$ required exploiting the modulation by Earth's rotation.
At the same time the present experiment provided an approach to the task of measuring
$\kappaezz$ with high accuracy (a weaker limit on this quantity can also be extracted from
 the experiment of Brillet and Hall).

Limitations of the experiment were the sensitivity of the optical path length
to temperature, and the limited laser lock quality as a consequence of the
relatively weak cavity throughput. This made the beat frequency more sensitive to optical path
length variations. In addition, a certain level of mechanical noise was present. 
In an upgraded experiment, obvious improvements are resonators of higher finesse and throughput, 
optical path length stabilization, and rotations with shorter period and lower tilt modulation.

In discussing a "null" experiment, it may be argued that a nonzero
value of the measured parameter may have been (partially) canceled
by an unknown  systematic effect, so that the bounds provided by an
experiment may be questioned. We emphasize the importance of the
fact that three experiments \cite{ant05,sta05,her05} have recently
been reported whose results are consistent with each other. Because
they were performed by independent groups with different techniques,
it is unlikely that they all exhibit a strong cancellation between
 the respective systematics and a substantial nonzero value of $\kappaezz$.

\vskip .2in
\noindent{\em Acknowledgments}
 {\small We thank E. G\"okl\"u for his participation in this work, L.~Haiberger for contributions to the cryostat development, A. Nevsky and C. L\"ammerzahl for discussions, 
and G. Thummes for his helpful assistance.
P.A. was supported by a DAAD fellowship, M.O. by a Heinrich-Hertz Foundation fellowship. 
This research was part of the Gerhard-Hess Program of the German Science Foundation.}

\small
\bibliographystyle{nature}
\bibliography{bibProceedings.bib}

\end{document}